\documentclass[11pt]{article}

\usepackage{amsmath,amssymb,latexsym,amsthm,epsfig,stmaryrd,tocbibind}

\def\part#1{\frac{\partial\phantom{q}}{\partial#1}}

\def\End{\mathop{\rm End}\nolimits}

\def\im{\mathop{\rm Im}\nolimits}

\def\cliff{\mbox{\sl Cliff}}

\def\mod{\mathop{\rm mod}\nolimits}
\def\re{\mathop{\rm Re}\nolimits}
\def\im{\mathop{\rm Im}\nolimits}

\newcommand{\be}[3]{\begin{equation}\label{#1#2#3}}
\newcommand{\bea}[3]{\begin{eqnarray}\label{#1#2#3}}
\newcommand{\ee}{\end{equation}}
\newcommand{\ba}{\begin{array}}
\newcommand{\ea}{\end{array}}
\newcommand{\eea}{\end{eqnarray}}
\newcommand{\norm}[1]{\parallel\!\!{#1}\!\!\parallel}

\newcommand{\htimes}{\widehat{\otimes}}
\newcommand{\R}{\mathbb{R}}
\newcommand{\C}{\mathbb{C}}

\newcommand{\mf}{\mathfrak}
\newcommand{\mc}{\mathcal}
\newcommand{\dstar}{d\!\star\!}
\newcommand{\haken}{\mathbin{\hbox to 8pt{\vrule height0.4pt width7pt depth0pt\kern-.4pt\vrule height4pt width0.4pt depth0pt\hss}}}

\newcommand{\mbf}{\mathbf}

\begin{document}

\thispagestyle{empty}

\begin{flushright}
\hfill{MPP-2004-170}\\
\hfill{hep-th/0412280}
\end{flushright}

\vspace{12pt}

\begin{center}{ \LARGE{\bf
Generalised $G_2$-structures \\[4mm]
and type IIB superstrings}}

\vspace{35pt}

{\bf Claus Jeschek}$^a$ and {\bf Frederik Witt}$^b$

\vspace{15pt}

$^a$ {\it  Max-Planck-Institut f\"ur Physik,\\
F\"ohringer Ring 6, 80805 M\"unchen, FRG\\
E-mail: jeschek@mppmu.mpg.de}\\[1mm]

\vspace{15pt}

$^b$ {\it  University of Oxford, Mathematical Institute,\\
24-29 St Giles, Oxford OX1 3LB, U.K.\\
E-mail: witt@maths.ox.ac.uk}\\[1mm]

\vspace{8pt}

\vspace{40pt}

{\bf ABSTRACT}

\end{center}
The recent mathematical literature introduces generalised geometries
which are defined by a reduction from the structure group $SO(d,d)$
of the vector bundle $T^d\oplus T^{d*}$ to a special subgroup. In
this article we show that compactification of IIB superstring vacua
on 7-manifolds with two covariantly constant spinors leads to a
generalised $G_2$-structure associated with a reduction from
$SO(7,7)$ to $G_2\times G_2$. We also consider compactifications on
6-manifolds where analogously we obtain a generalised
$SU(3)$-structure associated with $SU(3)\times SU(3)$, and show how
these relate to generalised $G_2$-structures.

\vfill

\newpage

\section{Introduction}

From a duality and a phenomenological point of view, the idea of
compactifying superstring theories and M-theory is a rather
appealing one. It also points to interesting geometrical issues as
requiring a certain amount of supersymmetry to be preserved puts
constraints on the internal background geometry and thus leads to
special $G$-structures. For instance, compactification on a 7- or
6-manifold together with $\mc{N}=1$ supersymmetry\footnote{Here,
$\mc{N}$ denotes the number of covariantly constant spinors in the
internal space.} yields a dilaton and a Killing spinor equation
which induces a $G_2$- or $SU(3)$-structure with various non-trivial
torsion classes.

For $\mc{N}=2$ supersymmetry we basically need two special
$G$-structures inside a given metric structure~\cite{gmw04}. This
naturally makes one consider the class of so-called {\em generalised
geometries}, a concept which goes back to~\cite{hi01}. Over a
$d$-manifold $M^d$, contraction on the vector bundle $T^d\oplus
T^{d*}$ defines an inner product of signature $(d,d)$ and therefore
induces an $SO(d,d)$-structure over $M^d$. A generalised geometry is
then described by the (topological) reduction to a special subgroup
of $SO(d,d)$ together with a suitable integrability condition. In
many situations, this can then be rephrased by the existence of two
$G$-structures with prescribed torsion classes and which sit inside
a given principal $SO(n)$- or $Spin(n)$-fibre bundle
(cf.~\cite{gu03},~\cite{wi04}).

In the present article we investigate type IIB superstring vacua by
compactifying on seven and six dimensional manifolds.\footnote{For
previous work see for
instance~\cite{gmpt04},~\cite{glmw02},~\cite{kmpt03}.} We shall
focus on the geometrical structure of the vacuum space admitting a
certain amount of supersymmetry in the external space, which can be
achieved by a spinorial formulation of the supersymmetry variations.
An additional analysis of the equations of motion single out
physical vacua.

As we will explain, considering the ``doubled" vector bundle
$T\oplus T^*$ accounts for $\mc{N}=2$ supersymmetry by reducing the
structure group from $SO(7,7)$ or $SO(6,6)$ to $G_2\times G_2$ or
$SU(3)\times SU(3)$ in the same way as compactification with
$\mc{N}=1$ supersymmetry is accounted for by a reduction from the
structure group $SO(7)$ or $SO(6)$ to $G_2$ or $SU(3)$. Our starting
point are the papers~\cite{ha00},~\cite{ha99},~\cite{has00}, where
Hassan investigated $T$-duality issues along $d$ directions. This
naturally leads to the consideration of ''general" supersymmetry
transformations invariant under the action of the so--called {\em
generalised T-duality} or {\em Narain group} $SO(d,d)$. Here,
``generalised" means that we treat the left- and right-moving
sectors of the worldsheet independently under the supersymmetry
variations. This is similar in spirit to the investigation of
topologically twisted non-linear sigma models on target spaces
admitting a generalised complex structure in the sense of
Hitchin~\cite{hi03} and Gualtieri~\cite{gu03}, see for instance
~\cite{cgj04},~\cite{gahuro84},~\cite{kali04},~\cite{ka03}.

By taking Hassan's supersymmetry variations, we see that preserving
supersymmetry on the external space requires the variations of the
gravitinos $\Psi_{\pm X}$ and dilatinos $\lambda_{\pm}$ to vanish,
i.e.
$$
\delta_{\pm}\Psi_{\pm \, X}  = 0 \quad\mbox{and}\quad
\delta_{\pm}\lambda_{\pm } = 0\,.
$$
These equations were also derived by Gauntlett et al.~\cite{gmpw04}
from the perspective of wrapped NS5-branes in IIB supergravity.
There, the authors found a solution by assuming ${\cal N}=2$
supersymmetry in dimension $d=3$ from a classical $SU(3)$-structure
point of view.

In this article we shall then proceed as follows. In
Section~\ref{prelim} we introduce the generalised supersymmetry
variations of type IIB following~\cite{ha00},~\cite{ha99}
and~\cite{has00}. Neglecting the action of the RR-sector we use
standard compactification methods in Section~\ref{compactification}
to obtain the model $\mathbb{R}^{1,2}\times M^7$ together with the
equations
\begin{equation}\label{dil_kill}
\Big(\gamma^a \partial_a \phi \mp {1 \over 12} \gamma^{abc} \,
H_{abc}\Big)\eta_{\pm}=0,\quad \Big( \partial_a +{1\over
4}(\omega_{abc}\mp {1 \over 2} H_{abc})\gamma^{bc}
\Big)\eta_{\pm}=0
\end{equation}
for two spinors $\eta_{\pm}$ on the internal background. In
section~\ref{geng2} we will discuss the notion of a generalised
$G_2$-structure~\cite{wi04} and give an equivalent formulation of
equations~(\ref{dil_kill}) by means of differential forms of mixed
degree. As a quick illustration we shall apply this formulation to
the setup taken from~\cite{gmpw04} in Section~\ref{su3}. We then
discuss compactifications on 6-manifolds which lead to generalised
$SU(3)$-structures in Section~\ref{gensu3}. Issues about lifting
string theory to generalised topological M-theory admitting a
generalised $G_2$-structure (see for instance~\cite{dgnv04})
naturally rises the question about the relation between generalised
$G_2$- and $SU(3)$-structures which will be addressed in
Section~\ref{genflow}.

The authors would like to acknowledge S.F.~Hassan and R.~Blumenhagen
for useful discussions. Some ideas are based on the second author's
doctoral thesis~\cite{wi05} who wishes to thank his supervisor Nigel
Hitchin.

\section{Preliminaries}\label{prelim}

We briefly set up the notation for the type IIB theory following
Hassan~\cite{ha00} (see also~\cite{ha99},~\cite{has00}). Let us
consider the gravitinos $\Psi_{\pm}$, the dilatinos $\lambda_{\pm}$
and the supersymmetry parameters $\epsilon_{\pm}$. They all arise as
the real and imaginary part of the complex Weyl spinors
$$
\Psi_X = \Psi_{+X} +i\Psi_{-X},\; X\in TM^{1,9}, \quad \lambda =
\lambda_+ +i\lambda_-\quad\mbox{and}\quad\epsilon=\epsilon_+
+i\epsilon_-.
$$
The gravitinos (of spin $3/2$) and the dilatinos (of spin $1/2$)
originate from the $(R,NS)$- and $(NS,R)$-sectors. We shall use an
additional subscript $+/-$ to indicate if the $R$-spin
representation is induced by the left or right moving sector. For
instance, $\Psi_{+X}$ comes from the $(R,NS)$-sector.

In the following we want to treat the left and right moving sector
independently with respect to space-time supersymmetry and we
therefore introduce two supersymmetry variations $\delta_{\pm}$. In
particular, $\delta_+$ and $\delta_-$ act only on the left and right
moving sectors and interchange $R\leftrightarrow NS$. Both the
supersymmetry parameters $\epsilon_{\pm}$ and the gravitinos
$\Psi_{\pm}$ are supposed to be of positive chirality, i.e.
$$
\Gamma^{11}\epsilon_{\pm}=\epsilon_{\pm} \qquad
\Gamma^{11}\Psi_{\pm}=\Psi_{\pm}
$$
while
$$
\Gamma^{11}\lambda_{\pm}=-\lambda_{\pm}\,,
$$
that is $\lambda_{\pm}$ is of negative chirality. We shall neglect
the action of the RR-fields and consider only the closed NS-NS
3-form flux $H$, i.e. $dH=0$, and the dilaton $\Phi$. Therefore the
supersymmetry variations of the $(R,NS)$- and $(NS,R)$-sector are
given by
\begin{equation}\label{susy}
\ba{rcl} \delta_{\pm}\Psi_{\pm \, X} & =& \nabla_X\epsilon_{\pm}
\mp\frac{1}{4} X\haken H\cdot\epsilon_{\pm},\; X\in TM^{1,9} \\
\delta_{\pm}\lambda_{\pm } & =& \frac{1}{2}( d\phi\mp\frac{1}{2}
H)\cdot\epsilon_{\pm}. \ea
\end{equation}
The vanishing of the supersymmetry variations, that is
\begin{equation}\label{susyvar}
\delta_{\pm}\Psi_{\pm \, X}  = 0\, , \quad\mbox{and}\quad
\delta_{\pm}\lambda_{\pm } = 0\, ,
\end{equation}
is necessary to characterise the background manifold $M^{1,9}$ in
the vacuum case. To find a solution to~(\ref{susyvar}), we shall
make specific assumptions which will occupy us next.

\section{Compactification on $M^7$}\label{compactification}

In this section, we compactify the theory on a 7-manifold $M^7$,
that is we consider the direct product model $\R^{1,2}\times M^7$
where $H$ and $\phi$ take non-trivial values only over $M^7$. We
want to determine the constraints on the underlying geometry of the
internal space $M^7$ imposed by the vanishing of the supersymmetry
variations~(\ref{susyvar}).

To that end, we decompose the supersymmetry parameters
$\epsilon_{\pm}\in \Delta^{\pm}_{M^{1,9}}$ accordingly, that is
$$
\epsilon_{\pm}=\sum_N \xi^N_{\pm}\otimes \eta^N_{\pm}\otimes{1
\choose 0},
$$
where $\xi$ and $\eta$ live in the irreducible spin representation
$\Delta_{\mathbb{R}^{1,2}}$ and $\mbf{8}$ of $Spin(1,2)$ and
$Spin(7)$ respectively, and $N\leq\mbox{dim}\,\mbf{8}=8$.

We fix the 10-dimensional space-time coordinates $X^M$
(M=0,\ldots,9) and assume the background fields to be independent of
$X^{\mu}$ ($\mu=0,1,2$). Coordinates on the internal space will be
labeled by $X^a$ for $a=3,\ldots,9$. We use the convention
$$
\{\Gamma^M,\Gamma^N\}=2\eta^{MN} \, \mathbb{I}_{32\times 32}
$$
with signature $(-,+,\ldots,+)$. We choose the explicit gamma matrix
representation
$$
\Gamma^M = \left\{
\begin{array}{r@{\quad:\quad}l}
\gamma_{\mu}\otimes\mathbb{I}_{8\times 8}\otimes\sigma_2& \mu=0,\ldots, 2\\
\mathbb{I}_{2\times 2}\otimes\gamma_a\otimes\sigma_1& a=3,\ldots, 9
\end{array}\right.
$$
where the $(8\times 8)$-matrices $\gamma_a$ are imaginary. The
$SO(1,2)$ gamma matrices $\gamma_{\mu}$ and the Pauli matrices
$\sigma_{i}$ are given by
$$
\gamma_0=\left(
\begin{array}{cc} 0 & -1 \\ 1 & 0 \end{array} \right) \qquad
\gamma_1=\left(
\begin{array}{cc} 0 & 1 \\ 1 & 0 \end{array} \right) \qquad
\gamma_2=\left( \begin{array}{cc} 1 & 0 \\ 0 & -1 \end{array}
\right)
$$
and
$$
\sigma_1=\left( \begin{array}{cc} 0 & 1 \\ 1 & 0
\end{array} \right) \qquad \sigma_2=\left( \begin{array}{cc} 0 & -i
\\ i & 0 \end{array} \right) \qquad \sigma_3=\left(
\begin{array}{cc} 1 & 0 \\ 0 & -1 \end{array} \right).
$$
Furthermore, we note the relations
$$
\prod_{\mu}\gamma_{\mu}= -\mathbb{I}_{2\times 2} \qquad
\prod_{a}\gamma_a=-i \,\mathbb{I}_{8\times 8}.
$$
The chirality operator $\Gamma^{11}$ is therefore $\Gamma^{11}=
\mathbb{I}_{2\times 2}\otimes\mathbb{I}_{8\times 8}\otimes
\sigma_3$.

With these splittings at hand we want to carry out the supersymmetry
variations~(\ref{susy}). The external part of the dilatino variation
vanishes trivially. For the internal part, we first note the useful
identity
$$
\Gamma^{M_1M_2M_3}H_{M_1M_2M_3}=(\mathbb{I}_{2\times
2}\otimes\gamma^{abc} \otimes\sigma_1) \, H_{abc}
$$
by means of which we immediately obtain the dilatino variations
$$
\delta_{\pm}\lambda_{\pm} = {1\over 2}\Big[\mathbb{I}_{2\times
2}\xi^N_{\pm} \otimes(\gamma^a \partial_a \phi \mp {1 \over 12}
\gamma^{abc} \, H_{abc})\eta^N_{\pm} \otimes \sigma_1 \, {1 \choose
0} \Big].
$$
The condition $\delta_{\pm}\lambda_{\pm} = 0$ is then equivalent to
\begin{equation}\label{dilaton}
\Big(\gamma^a \partial_a \phi \mp {1 \over 12} \gamma^{abc} \,
H_{abc}\Big)\eta^N_{\pm}=0.
\end{equation}
Next we focus on the variation of the gravitinos
$\delta_{\pm}\Psi_{\pm \, M}$. The flatness of $\R^{1,2}$ implies
$$
\nabla_{\mu}\xi^N_{\pm}=0.
$$
This solves the external part, and consequently we are left with
$$
\delta_{\pm}\Psi_{\pm \, a}  = \mathbb{I}_{2\times 2}\, \xi^N_{\pm}
\otimes\Big( \partial_a+{1\over 4}(\omega_{abc}\mp {1 \over 2}
H_{abc})\gamma^{bc} \Big)\eta^N_{\pm}\otimes {1 \choose 0}.
$$
Imposing the condition $\delta_{\pm}\Psi_{\pm \, X}=0$ finally
yields
\begin{equation}\label{genkill}
\Big( \partial_a +{1\over 4}(\omega_{abc}\mp {1 \over 2}
H_{abc})\gamma^{bc} \Big)\eta^N_{\pm}=0.
\end{equation}
In this article we shall deal with the case $N=1$, i.e. with exactly
two internal spinors $\eta_{\pm}$. Hence a solution consists of the
internal background data $(M^7,g,H,\eta_{\pm},\phi)$
satisfying~(\ref{dilaton}) and~(\ref{genkill}), where $g$ is a
metric, $H$ a closed 3-form, $\eta_{\pm}$ two unit spinors in the
associated irreducible spin representation $\mbf{8}$ and $\phi$ a
scalar function.

Note that the considerations above can be easily modified to tackle
the case of non-chiral type IIA theory which results in similar
geometric conditions.

\section{Generalised $G_2$-structures}\label{geng2}

In this section we want to formulate the geometry of the internal
space $M^7$ in the language of {\em generalised $G_2$-structures}
which we are going to define next. For details and proofs of the
facts below we refer to~\cite{wi04},~\cite{wi05} where these
structures where introduced.

Contraction on the vector bundle $T^d\oplus T^{d*}$ over an
arbitrary $d$-manifold defines a natural inner product of signature
$(d,d)$ and induces a spinnable $SO(d,d)$-structure. An element
$X\oplus\xi\in T^d\oplus T^{d*}$ acts on a form $\tau\in\Lambda^*$
by
$$
X\oplus\xi\bullet\tau=X\haken\tau+\xi\wedge\tau.
$$
As this squares to minus the identity,\footnote{We follow the usual
convention in mathematics where unit elements in the Clifford
algebra square to $-1$.} we obtain an isomorphism between
$\cliff(T^d\oplus T^{d*})$ and $\End(\Lambda^*)$. Moreover, the
irreducible spin representations of $Spin(d,d)$ can be realised as
$$
S^{\pm}=\Lambda^{ev,od}T^{d*}.
$$
In this way, an even or odd form $\rho$ may be regarded as a {\em
spinor} for $T^d\oplus T^{d*}$. In dimension 7, we call the pair
$(M^7,\rho)$ a {\em generalised $G_2$-manifold} if the stabiliser of
the even or odd form $\rho$ is conjugate to $G_2\times G_2$ inside
$Spin(7,7)$. Since a 2-form $b$ can be naturally identified with an
element in the Lie algebra $\mf{so}(d,d)$, it acts on $S^{\pm}$ by
wedging with the exponential
$e^b\bullet\tau=(1+b+b^2/2+\ldots)\wedge\tau$. Hence, if $\rho$
defines a generalised $G_2$-manifold, so does the transformed spinor
$e^b\wedge\rho$. This displays a crucial feature of generalised
geometries, namely that these can be naturally transformed by the
action of both diffeomorphisms and 2-forms (cf. also~\cite{hi03}
and~\cite{gu03}).

The data of the previous section links into the generalised setup as
follows. Consider a Riemannian 7-manifold $(M^7,g)$ together with
two unit spinors $\eta_+$ and $\eta_-$ living in the spinor bundle
associated with a fixed spin structure over $M^7$ and the
irreducible $Spin(7)$-representation $\mbf{8}$. In terms of
$G$-structures, it is a well-known fact that each spinor induces a
reduction to a principal $G_2$-fibre bundle inside this spin
structure. For sake of clarity, we denote the associated structure
groups by $G_{2\pm}$. The group $G_{2+}\times G_{2-}$ then fixes the
element $\eta_+\otimes\eta_-$ inside the irreducible $Spin(7)\times
Spin(7)$-representation space $\mbf{8}\otimes\mbf{8}$. On the other
hand, these groups also act on $\Lambda^{ev,od}$ via the inclusion
into $Spin(7,7)$. In order to compare these two actions we write
$$
\cliff(T^d,g)\htimes\cliff(T^d,-g)=\cliff(T^d\oplus T^{d*})
$$
where the isomorphism is given by extension of the map
$$
X\htimes Y\mapsto (X\oplus -X\haken g)\bullet(Y\oplus Y\haken g),
$$
and $\bullet$ now also denotes multiplication in $\cliff(T^d\oplus
T^{d*})$. Let $(\varphi\otimes\psi)^{ev,od}$ represent the even or
odd part of the form obtained through fierzing and let $\cdot$
denote Clifford multiplication in $\mbf{8}$. One can show that
\begin{equation}\label{lmap}
\begin{array}{rcl}
(X\cdot\varphi\otimes\psi)^{ev,od} & = &
X\htimes1\bullet(\varphi\otimes\psi)^{od,ev}\\
(\varphi\otimes Y\cdot\psi)^{ev,od} & = & \pm1\htimes
Y\bullet(\varphi\otimes\psi)^{od,ev}
\end{array}
\end{equation}
for any $\varphi,\,\psi\in\mbf{8}$. Hence the $G_{2+}\times
G_{2-}$-invariant tensor product $\eta_+\otimes\eta_-$ induces
elements $\rho_0^{ev,od}\in S^{\pm}$ whose stabiliser inside
$Spin(7,7)$ is conjugate to $G_2\times G_2$.

A particular instance of this construction was carried out
in~\cite{gmpw04} where the authors considered the case of two
orthogonal spinors $\eta_+$ and $\eta_-$, that is, the structure
group reduces to an honest $G_{2+}\cap G_{2-}=SU(3)$. In our
situation, the choice of the two spinors is perfectly general. To
fully appreciate that point, define the spinor
$\widetilde{\eta}_+=\eta_--q(\eta_-,\eta_+)\eta_+$ which is
orthogonal to $\eta_+$. It induces a vector field determined by the
relation $X\cdot\eta_+=\widetilde{\eta}_+$. Outside the zero locus
of $X$, the pair $(\eta_+,\widetilde{\eta}_+/\norm{X})$ induces an
$SU(3)$-structure on $M$ which breaks down precisely when $\eta_+$
and $\eta_-$ are parallel, that is, $G_{2+}\cap G_{2-}=G_2$. This is
also reflected in the following explicit description of
$\rho^{ev,od}_0$ in terms of the underlying $G_{2+}\cap
G_{2-}$-invariants. The coefficients of the form
$\eta_+\otimes\eta_-$ can be computed by
$$
g(\eta_+\otimes\eta_-,e_I)=q(e_I\cdot\eta_+,\eta_-),
$$
where $q$ denotes a suitably scaled $Spin(7)$-invariant inner
product on $\mbf{8}$ and $e_I=e_{i_1}\wedge\ldots\wedge e_{i_p}$ is
a multi-index of an orthonormal basis for $g$. Since $Spin(7)$ acts
transitively on the set of pairs of orthonormal spinors, we may
choose an orthonormal basis in $\mbf{8}$ such that
$$
\eta_+=(1,0,0,0,0,0,0,0,0)^{tr} \, , \quad\text{and}\quad
\eta_-=(\cos(a),\sin(a),0,0,0,0,0,0)^{tr}.
$$
If the spinors $\eta_+$ and $\eta_-$ are linearly independent, their
isotropy groups $G_{2+}$ and $G_{2-}$ intersect in $SU(3)$ which
fixes a 1-form $\alpha=e_7$, a symplectic form
$\omega=e_{12}+e_{34}+e_{56}$ and two 3-forms
$\psi_+=e_{135}-e_{146}-e_{236}-e_{245}$ and
$\psi_-=e_{136}+e_{145}+e_{235}-e_{246}$. These are the real and the
imaginary part of the $SU(3)$-invariant holomorphic volume
form~\cite{chsa02}. We then find
\begin{eqnarray}
\rho_0^{ev} & = & c+s\omega-c(\psi_-\wedge \alpha +
{\omega^2\over 2})+s\psi_+\wedge \alpha - s{\omega^3\over 6}\label{rhoeven}\\
\rho_0^{od} & = & s \alpha -c(\psi_+ +\omega\wedge \alpha)
-s\psi_--s{\omega^2\over 2}\wedge \alpha +c\, vol_g,\nonumber
\end{eqnarray}
where $c$ and $s$ are shorthand for $\cos(a)$ and $\sin(a)$, and
$a=\sphericalangle(\eta_+,\eta_-)$ describes the angle between the
spinors $\eta_+$ and $\eta_-$. The underlying $SU(3)$-structure
fluctuates with $a$ and breaks down when $s=0$, i.e. the spinors are
parallel. Consequently, only the forms $s\alpha$, $s\omega$ etc. are
globally defined over $M^7$ and it follows that in general the
$SO(7)$-structure does not reduce to a global ``static"
$SU(3)$-structure. Moreover, at a point where $a=0$, i.e.
$\eta=\eta_+=\eta_-$, we have
\begin{eqnarray*}
\rho^{ev} & = & \phantom{-}1-\star\varphi\\
\rho^{od} & = & -\varphi+vol,
\end{eqnarray*}
with $\varphi$ denoting the invariant 3-form of the $G_2$-structure
defined by $\eta$. This explicit description also reveals how to
relate the $G_2\times G_2$-invariant forms $\rho^{ev}_0$ and
$\rho^{od}_0$. For a $p$-form $\xi^p$, define $\sigma(\xi^p)$ to be
$1$ for $p\equiv0,3\mod \, 4$ and $-1$ for $p\equiv1,2\mod \, 4$. A
direct computation shows that
$$
\star\sigma(\eta_+\otimes\eta_-)^{ev,od}=(\eta_+\otimes\eta_-)^{od,ev}.
$$
In general, a $G_2\times G_2$-invariant spinor $\rho$ in $S^+$ or
$S^-$ determines a metric $g$ and a 2-form $b$ for which it can be
uniquely written (up to a sign) as
$$
\rho^{ev,od}=e^{-\phi} e^b\wedge(\eta_+\otimes\eta_-)^{ev,od}\in
S^{\pm}.
$$
Therefore, any generalised $G_2$-manifold can be equivalently
characterised by the set of data $(M^7,g,b,\eta_{\pm},\phi)$. We
refer to the induced 2-form $b$ as the {\em $B$-field} of the
generalised $G_2$-structure. In order to relate $\rho^{ev}$ with
$\rho^{od}$ we introduce the {\em generalised Hodge-} or {\em box
operator} $\Box_{g,b}:\Lambda^{ev,od}\to\Lambda^{od,ev}$ defined by
$$
\Box_{g,b}\rho^{ev,od} = e^b\wedge \star\sigma
(e^{-b}\wedge\rho^{ev,od}).
$$
One can then show that
$$
\Box_{\rho^{ev,od}}\rho^{ev,od}=\rho^{od,ev}.
$$
For sake of concreteness, we usually assume the $G_2\times
G_2$-invariant spinor $\rho$ to be even and write $\rho_0$ for the
$B$-field free form $\eta_+\otimes\eta_-$ given by~(\ref{rhoeven}).
We shall also use the sloppier notation $\Box_{\rho}$ for
$\Box_{g,b}$ if $g$ and $b$ are induced by $\rho$. Note that the
$$
g:\,28,\quad b:\,21,\quad\eta_+:\,7,\quad\eta_-:\,7,\quad\phi:\,1
$$
degrees of freedom sum to $64={\rm dim}\,\Lambda^{ev,od}$, so that
this data effectively parametrises the {\em open} orbit of a
$G_2\times G_2$-invariant form under the action of $\R_{>0}\times
Spin(7,7)$. Following the language in~\cite{hi01} such a spinor is
called {\em stable}. Stability allows us to consider a certain
variational principle introduced by Hitchin~\cite{hi01} which also
gained some attraction with a view towards a topological
M-theory~\cite{dgnv04},~\cite{pewi05}. The variation takes place over a
$d_H$-cohomology class, where $H$ is a closed 3-form and
$d_H=d+H\wedge$. If $\rho$ is $d_H$-closed, then it defines a
critical point for this variational problem if and only if
$d_H\Box_{\rho}\rho=0$. Note that if the spinors $\eta_+$ and
$\eta_-$ are equal and $H=0$, this equation reduces to the classical
condition $\dstar\varphi=0$~\cite{hi01}. To see how this relates
to~(\ref{dilaton}) and~(\ref{genkill}), recall that the twisted
Dirac operator over $\mbf{8}\otimes\mbf{8}$ transforms into
$d+d^{\star}$ under fierzing. A more general argument taking into
account the action of the 3-from $H$ can then be invoked to show
that $(M^7,g,H,\eta_{\pm},\phi)$ satisfies~(\ref{dilaton})
and~(\ref{genkill}) if and only if the corresponding $G_2\times
G_2$-invariant spinor $\rho_0=(\eta_+\otimes\eta_-)^{ev}$ satisfies
\begin{equation}\label{critpoint}
d_He^{-\phi}\rho_0=de^{-\phi}\rho_0+H\wedge e^{-\phi}\rho_0=0,\quad
d_H\Box_{\rho_0}e^{-\phi}\rho_0=d\Box_{\rho_0}e^{-\phi}\rho_0+H\wedge
\Box_{\rho_0}e^{-\phi}\rho_0=0,
\end{equation}
that is, $e^{-\phi}\rho_0$ defines a critical point. In particular,
we see that as a integrability condition we need
$$
e^{-\phi}\cos(a)=const.
$$
If $H$ is globally exact, i.e. $H=db$, (\ref{critpoint}) can be
written in the more succinct form
$$
de^{-\phi}(e^b\wedge\rho_0)=0,\quad
de^{-\phi}\Box_{g,b}(e^b\wedge\rho_0)=0.
$$

\section{Recovering the classical $SU(3)$-case}\label{su3}

Equations~(\ref{dilaton}) and~(\ref{genkill}) were first derived by
Gauntlett et al.~\cite{gmpw04} from a quite different point of view.
Starting with IIB supergravity they studied wrapped NS5-branes over
calibrated submanifolds inside an internal 7-manifold with an
$SU(3)$-structure. As an illustration of the previous section, we
reconsider their setup which turns out to be described by a
``static" generalised $G_2$-structure with $a\equiv\pi/2$ (that is,
the structure group reduces to a fixed $SU(3)$), together with an
additional closed 3-form, the NS-NS flux $H$.

Under this assumption the form $\rho$ defining the generalised
$G_2$-structure becomes
$$
\rho_0 = \omega +\psi_+\wedge\alpha-{\omega^3\over 6}
$$
with associated odd form
$$
\Box_{\rho_0}\rho_0 =  \alpha -\psi_--{\omega^2\over 2}\wedge\alpha.
$$
The supersymmetry equations are equivalent to
$$
d_He^{-\phi}\rho_0 = 0\quad\mbox{and}\quad
d_He^{-\phi}\Box_{\rho_0}\,\rho_0 = 0
$$
which written in homogeneous components can then be rephrased by
\begin{eqnarray*}
d\omega & = & d\phi\wedge\omega\,,\\
\psi_+\wedge d\alpha & = & -d\phi\wedge\psi_+\wedge \alpha
+d\psi_+\wedge\alpha - H\wedge\omega \,,\\
{1\over 2}d\omega\wedge\omega^2 & = & d\phi\wedge{\omega^3\over
6}-H\wedge\psi_+\wedge \alpha
\end{eqnarray*}
and
\begin{eqnarray*}
d\alpha & = & d\phi\wedge \alpha\,,\\
d\psi_- & = & d\phi\wedge\psi_- + H\wedge \alpha \,,\\
d\omega\wedge\omega\wedge\alpha & = & d\phi\wedge
\alpha\wedge{\omega^2\over2} - H\wedge\psi_-\,.
\end{eqnarray*}
We finally conclude
\begin{equation}
\begin{array}{rclcrcl}\label{finalres}
d(e^{-\phi}\alpha) & = & 0\,, & \quad & d(e^{-\phi}\omega) & = & 0\,,\\
\alpha\wedge d\psi_+ & = & H\wedge\omega\,, & \quad
& {1\over 3} d\omega\wedge\omega & = & \alpha\wedge H\wedge\psi_+\,,\\
d(e^{-\phi}\psi_-) & = & H\wedge\alpha\,, & \quad &
d\omega\wedge\omega\wedge\alpha & = & \psi_-\wedge H\,.
\end{array}
\end{equation}
The equations of motion are solved since $H$ is closed, i.e. $dH=0$,
as proved in~\cite{gmpw04}. Therefore~(\ref{finalres}) characterises
the physical vacua.

\section{Compactification on $M^6$ and generalised
$SU(3)$-structures}\label{gensu3}

Following the procedure of Section~\ref{compactification} we can
also compactify on a 6-dimensional manifold $M^6$. Recall that we
have $Spin(6)=SU(4)$ and that the irreducible spin representations
of positive and negative chirality $\Delta_{\pm}$ are just the
$SU(4)$-vector representation $\mbf{4}$ and its conjugate
$\overline{\mbf{4}}$. The supersymmetry equations compactified on
$M^6$ thus become
\begin{equation}\label{susy6dim}
\nabla^{LC}_X\eta_{\pm}\pm\frac{1}{4}X\haken
H\cdot\eta_{\pm}=0,\quad(d\phi\pm\frac{1}{2}H)\cdot\eta_{\pm}=0
\end{equation}
for two complex spinors $\eta_{\pm}$. Since we work in type IIB
theory both $\eta_+$ and $\eta_-$ are assumed to be of positive
chirality. Similarly, we can consider type IIA theory by choosing
the spinors to be non-chiral.

Recall that $SU(4)/SU(3)=S^7$, hence the choice of two unit spinors
$\eta_{\pm}\in\mbf{4}$ induces a reduction to two
$SU(3)_{\pm}$-subbundles. The $SU(4)$-representations $\Delta_{\pm}$
decompose into $\mbf{3}_{\pm}\oplus\mbf{1}_{\pm}$ and
$\overline{\mbf{3}}_{\pm}\oplus\overline{\mbf{1}}_{\pm}$.
Consequently, we can also consider the corresponding
$SU(3)_{\pm}$-invariant spinors
$\overline{\eta}_{\pm}\in\overline{\mbf{4}}$. We want to describe
the data $(M^6,g,H,\eta_{\pm},\phi)$ in the language of generalised
geometry where it gives rise to a generalised $SU(3)$-structure.
This is completely analogous to Section~\ref{geng2} and the proofs
of~\cite{wi04} carry over without difficulty. Again we content
ourselves with a brief outline of the corresponding results.

Rather than working with the complex spinors we will consider the
real $SU(4)$-module $S$ obtained by forgetting the complex structure
on $\mbf{4}$ or $\overline{\mbf{4}}$, that is the complexification
of $S$ is just $S^{\C}=\mbf{4}\oplus\overline{\mbf{4}}$. Note that
the Riemannian volume element $vol_g$ induces a complex structure on
$S$ and acts on $\mbf{4}$ and $\overline{\mbf{4}}$ by multiplication
with $i$ and $-i$ respectively. We let
$$
\varphi_{\pm}=\re(\eta_{\pm}),\quad
\widehat{\varphi}_{\pm}=\im(\eta_{\pm}),
$$
so that
$$
vol_g\cdot\varphi_{\pm}=-\widehat{\varphi}_{\pm}.
$$
Since $S$ carries an $SU(4)$-invariant Riemannian inner product, we
can identify $S\otimes S$ with $\Lambda^*T^{6*}$ through fierzing so
that~(\ref{lmap}) holds. This yields two forms
$\varphi_+\otimes\varphi_-$ and
$\varphi_+\otimes\widehat{\varphi}_-$ which we can interpret as
$SU(3)\times SU(3)$-invariant spinors and which we want to decompose
into an even and an odd part. Note that under complexification of
this isomorphism, the components $\mbf{4}\otimes\mbf{4}$ and
$\overline{\mbf{4}}\otimes\overline{\mbf{4}}$ get mapped onto odd
complex forms, while the off-diagonal components
$\overline{\mbf{4}}\otimes\mbf{4}$ and
$\mbf{4}\otimes\overline{\mbf{4}}$ become even since $\mbf{4}$ and
$\overline{\mbf{4}}$ are dual to each other. Writing
$\varphi_+\otimes\varphi_-=(\eta_++\overline{\eta}_+)\otimes(\eta_-+\overline{\eta}_-)/4$
etc. we obtain
$$
\begin{array}{l}
\tau_0=(\varphi_+\otimes\varphi_-)^{ev}=-(\widehat{\varphi}_+\otimes\widehat{\varphi}_-)^{ev}=\phantom{-}\frac{1}{2}\re(\eta_+\otimes\overline{\eta}_-)\\[5pt]
\widehat{\tau}_0=(\widehat{\varphi}_+\otimes\varphi_-)^{ev}=-(\varphi_+\otimes\widehat{\varphi}_-)^{ev}=\phantom{-}\frac{1}{2}\im(\eta_+\otimes\overline{\eta}_-)\\[5pt]
\upsilon_0=(\widehat{\varphi}_+\otimes\widehat{\varphi}_-)^{od}=-(\varphi_+\otimes\varphi_-)^{od}=-\frac{1}{2}\re(\eta_+\otimes\eta_-)\\[5pt]
\widehat{\upsilon}_0=(\varphi_+\otimes\widehat{\varphi}_-)^{od}=\phantom{-}(\widehat{\varphi}_+\otimes\varphi_-)^{od}=\phantom{-}\frac{1}{2}\im(\eta_+\otimes\eta_-).
\end{array}
$$
To see how these forms relate to each other, we note that in
dimension 6 the $\Box_{g,b}$-operator respects the parity of the
forms and satisfies $\Box_{g,b}^2=-Id$, that is $\Box_{g,b}$ induces
a complex structure on $\Lambda^*T^*$ (it is effectively the
$\wedge$-operator introduced in~\cite{hi03}). We then have
$$
\Box_{g,b}\tau_b=\widehat{\tau}_b,\quad\Box_{g,b}\upsilon_b=\widehat{\upsilon}_b,
$$
where $\tau_b=e^b\wedge\tau_0$ etc..

As in Section~\ref{geng2} we can compute a normal form description
which we can express in terms of the underlying $SU(2)=SU(3)_+\cap
SU(3)_-$-invariants if the unit spinors $\eta_{\pm}$ are linearly
independent. Using again the complexified isomorphism $S^{\C}\otimes
S^{\C}\cong\Lambda^*T^{6*\C}$ and decomposing
$\eta_-=c_1\eta_++c_2\eta_+^{\perp}$ with two complex scalars
$c_1,\,c_2\in\C$ we find
\begin{equation}\label{normform6dim1}
\eta_+\otimes\eta_-=i\bar{Z}\wedge(c_1\Omega+c_2e^{i\omega_1})
\end{equation}
and
\begin{equation}\label{normform6dim2}
\eta_+\otimes\overline{\eta}_-=e^{i\alpha\wedge\beta}\wedge(\bar{c}_1
e^{i\omega_1}+\bar{c}_2\Omega)
\end{equation}
where expressed in a suitable local orthonormal basis
$e_1,\ldots,e_6$ we have the two real 1-forms $\alpha=e_5$,
$\beta=e_6$, the complex 1-form $Z=e_5+ie_6$, the self-dual 2-forms
$\omega_1=e_{12}+e_{34}$, $\omega_2=e_{13}-e_{24}$,
$\omega_3=e_{14}+e_{23}$ and the complex symplectic form
$\Omega=\omega_2-i\omega_3$. The normal forms of
$\overline{\eta}_+\otimes\eta_-$ and
$\overline{\eta}_+\otimes\overline{\eta}_-$ are obtained by complex
conjugation in $\Lambda^*T^{6*\C}$.

Finally we wish to state the supersymmetry
equations~(\ref{susy6dim}) in terms of the $SU(3)\times
SU(3)$-invariant forms $\tau_0$, $\widehat{\tau}_0$, $\upsilon$ and
$\widehat{\upsilon}$. The real version of~(\ref{susy6dim}) is given
by
$$
\nabla^{LC}_X\varphi_{\pm}\pm\frac{1}{4}X\haken
H\cdot\varphi_{\pm}=0,\quad(d\phi\pm\frac{1}{2}H)\cdot\varphi_{\pm}=0
$$
and
$$
\nabla^{LC}_X\widehat{\varphi}_{\pm}\pm\frac{1}{4}X\haken
H\cdot\widehat{\varphi}_{\pm}=0,\quad(d\phi\pm\frac{1}{2}H)\cdot\widehat{\varphi}_{\pm}=0.
$$
The same computation as in the generalised $G_2$-case shows that
this is equivalent to
$$
d_He^{-\phi}\tau_0=d_He^{-\phi}\widehat{\tau}_0=0,\quad
d_He^{-\phi}\upsilon_0=d_He^{-\phi}\widehat{\upsilon}_0=0,
$$
that is
$$
d_He^{-\phi}\eta_+\otimes\eta_-=0,\quad
d_He^{-\phi}\overline{\eta}_+\otimes\eta_-=0.
$$
If $H$ is globally exact, that is $H=db$, we can write these
equations more succinctly as
$$
de^{-\phi}\tau_b=de^{-\phi}\Box_{g,b}\tau_b=0,\quad
de^{-\phi}\upsilon_b=de^{-\phi}\Box_{g,b}\upsilon_b=0.
$$

\section{Dimension 6 vs.\:7}\label{genflow}

The inclusion $SU(3)\subset G_2$ allows one to pass from an
$SU(3)$-structure in $\dim=6$ to a $G_2$-structure in $\dim=7$. In
the same vein, the inclusion $SU(3)\times SU(3)\subset G_2\times
G_2$ relates generalised $SU(3)$- to generalised $G_2$-structures.
In this final section we want to render this link explicit in both
the spinorial and the form picture of a generalised structure. We
first discuss the algebraic setup before we turn to integrability
issues.

To start with, assume that we are given a generalised
$G_2$-structure $(T,g,b,\eta_{\pm},\phi)$ over the 7-dimensional
vector space $T^7=T$ together with a preferred unit vector $\alpha$.
We want to induce a generalised $SU(3)$-structure on
$T^6=\widetilde{T}$ defined by $T=\widetilde{T}\oplus\R\alpha$.
Since $\alpha\cdot\alpha=-1$, the choice of such a vector induces a
complex structure on the irreducible $Spin(7)$-module $\mbf{8}$
which is compatible with the spin-invariant Riemannian inner
product. Hence the complexification of $\mbf{8}$ is
$$
\mbf{8}\otimes\C=\Delta^{1,0}\oplus\Delta^{0,1},
$$
where
$$
\Delta^{1,0/0,1}=\{\eta\mp i\alpha\cdot\eta\:|\:\eta\in\Delta\}.
$$
The choice of $\alpha$ also induces a reduction from $SO(7)$ to
$SO(6)$ which is covered by $Spin(6)=SU(4)$, and as an
$SU(4)$-module we have $\Delta^{1,0}=\mbf{4}$ and
$\Delta^{0,1}=\overline{\mbf{4}}$. We define
$$
\psi_{\pm}=\eta_{\pm}- i\alpha\cdot\eta_{\pm}
$$
and let $\widetilde{g}=g_{|\widetilde{T}}$ and
$\widetilde{b}=\alpha\haken(\alpha\wedge b)$. Then a generalised
$SU(3)$-structure over $\widetilde{T}$ is given by
$(\widetilde{T},\widetilde{g},\widetilde{b},\psi_{\pm},\phi)$.
Moreover, we get a (possibly zero) 1-form $\beta=\alpha\haken
b\in\Lambda^1\widetilde{T}^*$. It is clear that we can reverse this
construction by defining a metric
$g=\widetilde{g}+\alpha\otimes\alpha$,
$b=\widetilde{b}+\alpha\wedge\beta$ and two spinors
$\eta_{\pm}\in\mbf{8}$ through $\eta_{\pm}=\re(\psi_{\pm})$.

To see what happens in the form picture, we start with the special
$G_2\times G_2$-invariant form
$$
\rho_0=(\eta_+\otimes\eta_-)^{ev}=f_0+\alpha\wedge f_1,
$$
where $f_0\in\Lambda^{ev}\widetilde{T}^*$ and
$f_1\in\Lambda^{od}\widetilde{T}^*$. It follows from~(\ref{lmap})
that
\begin{eqnarray*}
\alpha\wedge(\eta_+\otimes\eta_-)^{ev,od} & = & \frac{1}{2}(\alpha\cdot\eta_+\otimes\eta_-\mp\eta_+\otimes\alpha\cdot\eta_-)^{od,ev}\\
\alpha\haken(\eta_+\otimes\eta_-)^{ev,od} & = &
\frac{1}{2}(-\alpha\cdot\eta_+\otimes\eta_-\mp\eta_+\otimes\alpha\cdot\eta_-)^{od,ev}.\\
\end{eqnarray*}
Therefore the forms $f_0$ and $f_1$ can be expressed by
$$
f_0
=\alpha\haken(\alpha\wedge\rho_0)=\frac{1}{2}(\eta_+\otimes\eta_-+\alpha\cdot\eta_+\otimes
\alpha\cdot\eta_-)^{ev}
$$
and
$$
f_1=\alpha\haken\rho_0=-\frac{1}{2}(\alpha\cdot\eta_+\otimes\eta_-+
\eta_+\otimes \alpha\cdot\eta_-)^{od}.
$$
Using the spinors $\psi_{\pm}$ as defined above we find
\begin{eqnarray*}
\psi_+\otimes\psi_- & = & (\eta_+\otimes\eta_- -
\alpha\cdot\eta_+\otimes \alpha\cdot\eta_-)- i  (\alpha\cdot\eta_+\otimes\eta_- + \eta_+\otimes \alpha\cdot \eta_-)\\
\psi_+\otimes\overline{\psi}_- & = & (\eta_+\otimes\eta_- +
\alpha\cdot\eta_+\otimes \alpha\cdot\eta_-)+ i  (- \alpha\cdot\eta_+\otimes\eta_-+\eta_+\otimes \alpha\cdot\eta_- ).\\
\end{eqnarray*}
Consequently, we have
$$
f_0=\frac{1}{2}\re(\psi_+\otimes\overline{\psi}_-)=\tau_0
$$
and
$$
f_1= \frac{1}{2}\im(\psi_+\otimes\psi_-)=\widehat{\upsilon}_0.
$$
In the same vein, decomposing $\Box_{g,0}\rho_0=g_1+\alpha\wedge
g_0$ yields
$$
g_0=\frac{1}{2}\im(\psi_+\otimes\overline{\psi}_-)=\widehat{\tau}_0
$$
and
$$
g_1=\frac{1}{2}\re(\psi_+\otimes\psi_-)=-\upsilon_0.
$$
In presence of a non-trivial $B$-field $b\in\Lambda^2T^*$ we write
$b=\widetilde{b}+\alpha\wedge\beta$. Since
$e^{\widetilde{b}+\alpha\wedge\beta}=e^{\widetilde{b}}\wedge(1+\alpha\wedge\beta)$
we obtain for the general case the expressions
\begin{equation}\label{g2su3rho}
\rho_b=e^{-\phi}\tau_{\widetilde{b}}+\alpha\wedge(e^{-\phi}\widehat{\upsilon}_{\widetilde{b}}+\beta\wedge
e^{-\phi}\tau_{\widetilde{b}})
\end{equation}
and
\begin{equation}\label{g2su3boxrho}
\Box_{g,b}\rho=-e^{-\phi}\upsilon_{\widetilde{b}}+\alpha\wedge(e^{-\phi}\widehat{\tau}_{\widetilde{b}}-\beta\wedge
e^{-\phi}\upsilon_{\widetilde{b}}).
\end{equation}
Conversely, if $(T,g,b,\rho_0,\phi)$ defines a generalised
$G_2$-structure and $\alpha\in T$ is a unit vector, then the forms
$\widetilde{b}=\alpha\haken b$,
$\tau_0=\alpha\haken(\alpha\wedge\rho_0)$ and
$\upsilon_0=-\alpha\haken\Box_{g,0}\rho_0$ define a generalised
$SU(3)$-structure
$(\widetilde{T},\widetilde{g},\widetilde{b},\tau_0,\upsilon_0,\phi)$
with $\widetilde{g}=g_{|T}$.

To see how the integrability conditions relate to each other over
the manifolds $M^7=M$ and $M^6=\widetilde{M}$, consider a smooth
family
$(\widetilde{g}(t),\widetilde{b}(t),\tau_0(t),\upsilon_0(t),\phi(t))$
of metrics $\widetilde{g}(t)$, of 2-forms $\widetilde{b}(t)$, of
even and odd forms $\tau_0(t)$ and $\upsilon_0(t)$ and of scalar
functions $\phi(t)$ which we assume to define a generalised
$SU(3)$-structure for any $t$ lying in some open interval $I$.
Moreover, we consider a curve of 1-forms
$\beta(t)\in\Omega^1(\widetilde{M})$. In order to obtain an
integrable generalised $G_2$-structure over $\widetilde{M}\times I$
defined by $(\widetilde{M}\times I,g,b,\rho_0,\phi)$ where
$g=\widetilde{g}_t\oplus dt\otimes dt$,
$b=\widetilde{b}(t)+dt\wedge\beta(t)$,
$\rho_0=\tau_0(t)+dt\wedge\widehat{\upsilon}_0(t)$ and
$\phi=\phi(t)$, we need to solve the equations
$$
d\rho=0,\quad d\Box_{g,b}\rho=0.
$$
We decompose the exterior differential $d$ over
$M=\widetilde{M}\times I$ into
$$
d_{|\Omega^{ev,od}}\cdot \rightarrow
d_{|\Omega^{ev,od}}\cdot=\widetilde{d}_{|\Omega^{ev,od}}\cdot\pm\partial_{t|\Omega^{ev,od}}\cdot\wedge
dt,
$$
where $\widetilde{d}$ is the exterior differential on
$\widetilde{M}$. From~(\ref{g2su3rho}) we conclude the first
equation to be equivalent to
$$
d\rho=\widetilde{d}e^{-\phi}\tau_{\widetilde{b}}+dt\wedge(\partial_te^{-\phi}\tau_{\widetilde{b}}-\widetilde{d}e^{-\phi}\widehat{\upsilon}_{\widetilde{b}}-\widetilde{d}(\beta\wedge
e^{-\phi}\tau_{\widetilde{b}}))
$$
so that
\begin{equation}\label{cond1}
\widetilde{d}e^{-\phi}\tau_{\widetilde{b}}=0,\quad
\partial_te^{-\phi}\tau_{\widetilde{b}}=\widetilde{d}e^{-\phi}\widehat{\upsilon}_{\widetilde{b}}+\widetilde{d}\beta\wedge
e^{-\phi}\tau_{\widetilde{b}}.
\end{equation}
By~(\ref{g2su3boxrho}) the second equation reads
$$
d\Box\rho=-\widetilde{d}e^{-\phi}\tau_{\widetilde{b}}+dt\wedge(-\partial_te^{-\phi}\upsilon_{\widetilde{b}}-\widetilde{d}e^{-\phi}\widehat{\tau}_{\widetilde{b}}-\widetilde{d}(\beta\wedge
e^{-\phi}\upsilon_{\widetilde{b}}))
$$
and therefore yields
\begin{equation}\label{cond2}
\widetilde{d}e^{-\phi}\upsilon_{\widetilde{b}}=0,\quad
\partial_te^{-\phi}\upsilon_{\widetilde{b}}=-\widetilde{d}e^{-\phi}\widehat{\tau}_{\widetilde{b}}+\widetilde{d}\beta\wedge
e^{-\phi}\upsilon_{\widetilde{b}}.
\end{equation}
If we let $\bar{\beta}(t)=\int_0^s\beta(s)ds$ we can
bring~(\ref{cond1}) and~(\ref{cond2}) into Hamiltonian form, that is
the generalised $G_2$-structure is integrable if and only if
\begin{equation}\label{flow}
\begin{array}{rclcrcl}
\widetilde{d}(e^{-\phi}e^{\widetilde{d}\bar{\beta}}\wedge\upsilon_{\widetilde{b}})
& = & 0\,,& \quad & \widetilde{d}\widehat{\upsilon} & = &
\partial_t(e^{-\phi}e^{\widetilde{d}\bar{\beta}}\wedge\tau_{\widetilde{b}})\,,\\
\widetilde{d}(e^{-\phi}e^{\widetilde{d}\bar{\beta}}\wedge\tau_{\widetilde{b}})
& = & 0\,,& \quad & \widetilde{d}\widehat{\tau} & = &
-\partial_t(e^{-\phi}e^{\widetilde{d}\bar{\beta}}\wedge\upsilon_{\widetilde{b}})\,.
\end{array}
\end{equation}
We illustrate the previous discussion by considering a classical
$SU(3)$-structure defined by a unit spinor $\eta$ and taking the
generalised $SU(3)$-structure given by $(M^6,g,\eta)$ with trivial
B-field and vanishing dilaton, i.e. $b=0$ and $\phi=0$. We can
compute the forms $\tau_0$ etc. by using the normal form
description~(\ref{normform6dim1}) and~(\ref{normform6dim2}) where
$c_1=1$ and $c_2=0$. Using the notation of~\cite{chsa02} and
Section~\ref{geng2} we obtain
$$
\begin{array}{rclcrcl}
\upsilon_0 &=&  -{1\over 2}\psi_-\,, &\quad
            & \widehat{\upsilon}_0 & = &{1\over 2}\psi_+\,,\\
\tau_0 & = & {1\over 2}(1- {\omega^2\over 2})\,, &\quad
            & \widehat{\tau}_0 &= &{1\over 2}(\omega- {\omega^3\over 6})\,,
\end{array}
$$
and equations~(\ref{flow}) become the Hitchin flow equations
$$
\begin{array}{rclcrcl}
\widetilde{d}\psi_- & = & 0\,, & \quad
    &\widetilde{d}\psi_+ & = & -\partial_t\omega\wedge\omega\,,\\
\widetilde{d}\omega\wedge\omega & = & 0\,, &\quad
   &\widetilde{d}\omega & = & \phantom{-}\partial_t\psi_-\,,
\end{array}
$$
which appeared in~\cite{chsa02} and go back to~\cite{hi01}. Note
that although equations~(\ref{flow}) are, like the Hitchin flow
equations, in Hamiltonian form we have not shown yet that if the
data
$(\widetilde{g}(t),\widetilde{b}(t),\tau_0(t),\upsilon_0(t),\phi(t))$
defines a generalised $SU(3)$-structure at $t=t_0$ and
satisfies~(\ref{flow}), then it automatically defines a generalised
$SU(3)$-structure for $t>t_0$, as it is the case for classical
$SU(3)$-structures evolving along the Hitchin flow.

\bibliographystyle{gillow}

\end{document}